\documentclass[aps,pra,twocolumn,showpacs,superscriptaddress,groupedaddress]{revtex4-1}
\usepackage[margin={2cm,2cm}]{geometry}
\usepackage{graphicx}
\usepackage{natbib}
\usepackage{amsthm}
\usepackage[english]{babel}

\usepackage{amsmath}
\usepackage{amsfonts}
\usepackage{amssymb}
\usepackage{subfig}
\usepackage{hyperref}
\usepackage[usenames, dvipsnames]{color}
\usepackage{bbold}
\usepackage{bbm}
\usepackage{amsfonts}
\usepackage{hyperref}
\usepackage[english]{babel}
\hypersetup{
    colorlinks=true,
    linkcolor=ForestGreen,
    filecolor=magenta,      
    urlcolor=cyan,
    citecolor=blue
}
\usepackage{epstopdf}
\usepackage{lipsum}

\begin{document}
 \title{A generalised framework for non-classicality of states}
  \author{Soumik Adhikary}
  \email{soumikadhikary@physics.iitd.ac.in}
  \affiliation{Department of Physics, Indian Institute of Technology Delhi, New Delhi-110016, India.}

 \author{Sooryansh Asthana}
  \email{sooryansh.asthana@physics.iitd.ac.in}
  \affiliation{Department of Physics, Indian Institute of Technology Delhi, New Delhi-110016, India.}

  \author{V. Ravishankar}
  \email{vravi@physics.iitd.ac.in}
  \affiliation{Department of Physics, Indian Institute of Technology Delhi, New Delhi-110016, India.}
  
\begin{abstract} Non-classical  probability  (along with its underlying logic) is a   defining feature of quantum mechanics. A formulation that incorporates them, inherently and directly,  would promise  a unified description of seemingly different prescriptions of non-classicality of states that have been proposed so far. This paper sets up such a formalism. It is based on elementary considerations, free of ad-hoc definitions,  and is completely operational. It permits a systematic construction of non-classicality conditions on states and also to quantify the non-classicality, at the same time. This quantification, as shown for the example of two level systems, can serve as a measure of coherence and can be furthermore, harnessed to obtain a measure for pure state entanglement for coupled two level systems.

  \end{abstract}
  

\maketitle  

\section{Introduction}

Quantum mechanics 
has altered the very way we comprehend  laws of nature. Equally so, it has altered  the way we formulate  laws of probability.   Thus, the concept of non-classicality of states in quantum physics   is as much a reflection of the new    probability  as it is  of non-classical physics. This was recognised quite early, in a rather formal manner, by Birkhoff and von Neumann  \cite{birkhoff36} (see also \cite{jauch69,Accardi81}). Recent developments in quantum information, have brought the  realization that  non-classicality of quantum states   can, in fact, act as  resources  for information processing,  some of which could even   be impossible otherwise\cite{Benn84,Benn93}.  
In consequence, many definitions and criteria have been proposed \cite{bellinequalities, CHSHOrig, BellSpeakable, Kochen67, Werner89, Horodecki96b, Woot98, Coll02, Olliv01, Wiseman07,Alicki08,Adhikary16}.  In parallel, 
there has been a vigorous experimental activity, both for probing the foundations of quantum mechanics\cite{Aspect81,Aspect82,Shalm15,Guis15}  and for eminently practical applications\cite{Benn84,Shor97,Leach09,Ren17}.

Spectacular though these developments are, our  present  understanding of non-classicality is not entirely satisfactory. Each  definition/criterion is pinned to a specific context, and its  interrelationship with other criteria  is not always clear.  It is, therefore,  highly desirable to formulate nonclassicality directly in the language of quantum probabilities and their underlying logic. This would require setting up of a formalism
 which has (i) non-classical logic (in the sense of \cite{birkhoff36}) and probability inbuilt in it, (ii) is completely operational and which, in particular, allows for a systematic study    of non-classicality conditions. Finally, it should be free of ad-hoc constructions.
  This task, if accomplished,  would provide  a unified framework to describe non-classicality of states  and allow a systematic way for devising tests for verifying non-classicality of any given state. 
 
  This paper sets up such a formalism.  Starting from well established, simple, but equally general rules of quantum mechanics, we introduce  what we call pseudo projections, and
   show how they may be harnessed to obtain  an  infinitely large number of tests of non-classicality of states, within a single framework. This formulation describes  non-classicality in a broad setting and hence serves as an encompassing framework. Being operational it also automatically `quantifies' non-classicality. Keeping this operational perspective in mind, we do not wade into the more complicated and unresolved issues concerning the logical foundations of  quantum mechanics or of quantum probability, beyond making some essential observations in Section  \ref{s2e}.

 After setting up the formalism we illustrate it's applicability for the simple case of a qubit system. As yet another preliminary application, we show that the quantification of non-classicality through this framework, yields an entanglement monotone for pure two qubit states.

\section{The Formalism}
 It is convenient to  start with  a  query,  articulated  clearly  by Fine\cite{Fine82, Fine82-2}: are there circumstances under which a given quantum state permits assignments of joint probabilities for the outcomes of a given set of  incompatible observables?  This important question, which  forms the basis of our analysis,  finds its  mathematical expression   in pseudo-projections.

\subsection{Joint probabilities, conjunctions and pseudo-projections}

Consider a set of  observables, $A^1, A^2, \dots, A^r$, defined over a  phase space ${\Phi}$. Let
  $A^k$ take values belonging to a set $\{a^k_1, a^k_2,  \cdots\}$. For a classical system  in a  state $f$,  the joint probability for a  conjunction of events $\{ A^1 = a^1_{i_1}, A^2 = a^2_{i_2}, \cdots, A^r = a^r_{i_r}\}$  always exists and admits a  straightforward construction: let $S \subset \Phi$ be the support for the joint outcomes, and   $\mathbf{1}_S $, its indicator function. Then,  the joint probability is given by the overlap of $f$ with $\mathbf{1}_S$. By itself, $\mathbf{1}_S$ is  a Boolean  observable: it takes  value 1 in $ S$ and vanishes outside the support.  If $S_{i_{\alpha}}$ are supports for individual outcomes, then $S =\bigcap\limits_{i_{\alpha}}^{}S_{i_{\alpha}}$, which is the set theoretical representation of conjunction.

  Observables in quantum mechanics  are represented by  hermitian operators 
 in a Hilbert space ${\cal H}$. 
 Suppose that   $A = \sum_i a_i \pi_i$ is the eigen-resolution of an observable $A$. The eigen-projections $\pi_i$ are the quantum representatives of  indicator functions;  the eigen-projections partition the Hilbert space into a disjoint direct sum of eigenspaces,  ${\cal H} = \bigcup\limits_{i}^{}\oplus{\cal H}_i$. The  subspace ${\cal H}_i \equiv \pi_i {\cal H}$ is  the representative of the corresponding support $S_i$.  The probability for the outcome $a_i$, for a state $\rho$, is given by the 
 overlap, $Tr(\rho\pi_i)$.  Unless the spectrum is continuous, most indicator functions would map to the trivial null projection, of rank zero.  The indicator function with the full phase space as its support maps to the identity operator, $\mathbb{1}$.
 
\begin{figure}  
\centering       
\includegraphics[width = 0.5 \textwidth]{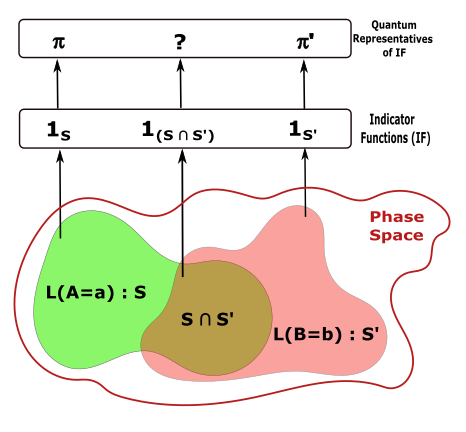}
\centering
\caption{\label{schematic} A schematic diagram depicting the indicator functions and their quantum representatives for a pair of events and their conjunction.}
\end{figure}

 The crux of the matter is that indicator functions for joint outcomes of observables do not necessarily map to   projection operators. This is the statement of uncertainty principle in its primordial form.  Nevertheless,  we may still inquire into  the quantum  representatives -  hermitian operators -  of such indicator functions. This is depicted  in  Fig.~\ref{schematic} schematically. The quantum representatives will be called  pseudo-projections.  Pseudo-projections hold the key to non-classicality. 

\subsubsection{Construction of Pseudo-projections}   
\label{cps}
  We follow standard rules of quantum mechanics for construction of pseudo-projections.  
  Consider, first,  two observables $A,B$. Let $\pi^A_{a_i},~\pi^B_{b_j}$
be the projection operators representing the respective indicator functions  for the outcomes $A = a_i$ and $B = b_j$.  The unique operator, the {\it  pseudo-projection},  that represents the indicator function,
$\mathbf{1}_{S_i \cap S_j}$,
 for the corresponding classical joint outcome,  is given by the symmetrised product:
 \begin{equation}
 \mathbf{\Pi}^{AB}_{a_i b_j} = \frac{1}{2}\{ \pi^A_{a_i},\pi^B_{b_j} \}, 
 \end{equation}
 which is but the simplest example of Weyl ordering \cite{Weyl27}.  Though  it is hermitian, it is not idempotent  and, in fact, admits negative eigenvalues,unless $[\pi^A_{a_i},~\pi^B_{b_j}]=0$, i.e., $\mathbf{\Pi}^{AB}_{a_i b_j}$ is itself a projection. This important property is proved in the appendix.  The  pseudo projection $\mathbf{\Pi}^{AB}_{a_i b_j}$  vanishes identically if $\pi^A_{a_i}= \mathbb{1} - \pi^B_{b_j}$. This property is consistent with the requirement that the joint probability for an event and its negation should vanish identically \footnote{More sophisticated ramifications involving multiple observables will be discussed separately}.

Pseudo-projections that represent  an indicator function for joint outcomes of more  than two observables are not unique, reflecting the inherent ambiguity in constructing quantum representatives of classical observables. Symbolically, let ${\Pi}^N_{\{\alpha\}}$ be a product of  $N$  projections, in some order. The hermitian combination, $\frac{1}{2}(\Pi^N_{\{\alpha\}} + \Pi_{\{\alpha\}}^{N\dagger})$,  is a legitimate representative  of the  indicator function  for the corresponding joint event.  We call this  a unit pseudo-projection. For each such classical event there are, in general, $\frac{N!}{2}$ unit projections, which are distinct if no two projection operators commute. Should all the projections corresponding to the $N$ outcomes commute, the manifold of pseudo-projections collapses to a single point and represents a true projection. The convex linear span of distinct unit pseudo-projections yields the family of pseudo-projections that  represent the parent indicator function. 
This fact makes the discussion of joint probability for quantum states richer. Though each choice is equally admissible, and leads to negative eigenvalue(s) in the representative pseudo projections, the one obtained from Weyl ordering, which is the sum of all unit pseudo-projections with equal weights, seems the most favoured: it is completely symmetric in all the projections, as its classical counterpart is, and  has a close relationship with Moyal brackets \cite{Moyal49} and Wigner distribution functions \cite{Wigner32} which are central to semi classical descriptions.

\subsection{Pseudo-projections and pseudo-probabilities}

 Pseudo-projections generate pseudo-probabilities. 
Let $\mathbf{\Pi}^N(\{A^{\alpha} = a^{\alpha}_{i_{\alpha}}\});~\alpha =1,\cdots,N$, denote the pseudo-projection for the set of the outcomes of $N$ observables indicated in the parenthesis.  For a system in a state $\rho$,  it generates a  pseudo--probability which may be defined as the expectation
\begin{equation}\label{pp}
{\cal P}(\{ A^{\alpha} = a^{\alpha}_{i_{\alpha}}\} )  \doteq 
  Tr\Big[\rho \mathbf{\Pi}^N(\{A^{\alpha} = a^{\alpha}_{i_{\alpha}}\})\Big].
\end{equation}
 Pseudo-probabilities can take negative values. 
 A complete set of pseudo-probabilities, corresponding to all possible joint outcomes of a given set of observables, 
 is the quantum analogue of the classical joint probability scheme.  We call this
 the \textit{ pseudo-probability scheme}, or in short,  the scheme, henceforth.  The entries in the schemes do add to unity. The  marginal schemes, corresponding to   sets of  mutually commuting observables,  are always  true probability schemes, and have entries as mandated by quantum mechanics.
 
Pseudo-probability schemes serve to define non-classicality of states in an encompassing  manner. 

\subsection{Non-classicality of states} 
{\bf Definition:} {\it Let ${\cal S}_{\rho}(\{A^{\alpha}\})$ be the  scheme for a set of $N$ observables  for a system in a  state $\rho$. 
The  state is  non-classical with respect to these observables if, even one of the pseudo-probabilities in the scheme is  negative. It  may be deemed to be classical  if, and only if, all the entries  are non-negative.}\\

If no restriction is placed on the observables, all states, without exception,  turn out to be  quantum, as we show in  section~\ref{threeobv}.  However, in importance and as  resources, not all states are on equal footing. Therefore,  if reasonable restrictions are placed on the choice of observables, or 
if only certain  combinations of pseudo-probabilities are forced, 
some states would be classical, and others -- non-classical. The many prescriptions for designating  states as non-classical  e.g, \cite{bellinequalities,CHSHOrig, Olliv01}  belong to this category. In short, classicality of a state is relative and not absolute. 

  Suppose that  all the entries in a pseudo-probability scheme are non-negative, even when the underlying projections are mutually non-commuting.   It would mean, as it were, that the scheme represents  joint probabilities  for  a correlated classical state, with quantum probabilities as its marginals.     Such a   scheme  automatically  yields an explicit construction of contextual hidden parameters for those non-commuting observables\cite{Klyachko08}. 
  It is important to note that the  hidden parameters or, equivalently, the underlying classical systems are not necessarily
  defined over the same classical phase space that one starts with. This fact becomes clear in our discussion in Section \ref{s3}.
  
    Also, along the sidelines,  the definition brings out the 
 real import of the idea of negative probability advocated by 
 Dirac,\cite{Dirac42}, Bartlett  \cite{Bartlett45}, and most forcefully, by Feynman \cite{Feynman87}. It has been argued that the introduction of negative probability, even if it be in an {\it ad hoc} manner (as in p 11 of   \cite{Feynman87}),  does have a role in the sense of consistent book keeping in intermediate processes (and calculations).  This intuitive idea,  set forth in \cite{Dirac42,Feynman87},  gets automatically incorporated in the present formalism.  The smallest subset in the event space for which the pseudo-probability is not negative would give the minimum coarse graining required for a physical interpretation of joint probability. For example, for pairs of observables  $\sum_{b_j} {\cal P}_{a_i b_j}$ always gives a physically realisable scheme with non-negative entries,
 but there could be partial sums which yield the same.

\subsection{Disjunction and negation}
Complete information on non-classicality is captured by  schemes, which correspond to the logical conjunction(AND). The quantum representatives of the indicator function corresponding to the two other operations, disjunction (OR) and negation (NOT),
may be obtained from their appropriate parent scheme by  applying standard probability rules. 
Thus, the quantum representative of the indicator function representing the disjunction,  $A= a_i$ OR $B=b_j$, which we shall denote ${\bf R}^{AB}_{a_ib_j}$,  has the expression:
\begin{eqnarray}
\label{ORproj}
{\bf R}^{AB}_{a_ib_j} &=& \sum_k {\bf \Pi}^{AB}_{a_ib_k} + \sum_k {\bf \Pi}^{AB}_{a_kb_j} - {\bf \Pi}^{AB}_{a_ib_j} \\ \nonumber
&=& \pi^A_{a_i} + \pi^B_{b_j} - {\bf \Pi}^{AB}_{a_ib_j}.
\end{eqnarray}
In a similar manner, the negation of a pseudo-projection is represented by subtracting it from the identity \footnote{This formal definition, however, fails to obey the fundamental classical rule that would require ${\bf N}^{AB}_{a_ib_j} {\bf \Pi}^{AB}_{a_ib_j} = 0$.}
\begin{equation}
\label{NOTproj}
{\bf N}^{AB}_{a_ib_j} = \mathbb{1} - {\bf \Pi}^{AB}_{a_ib_j}.
\end{equation}

\subsection{Differences with   classical logic and Kolmogorov formulation}
\label{s2e}
We conclude this section by mentioning briefly how quantum logic (probability) deviates from the classical (Kolmogorov) formulation. First of all, consider pseudo projections. Non-classical logic, as understood in \cite{birkhoff36},  is inherent in the very  definition of pseudo projections. 
  The formalism developed in \cite{birkhoff36} identifies joint outcomes of non-commuting observables with absurd propositions in the quantum domain, although they are perfectly valid classically. The violations of  Boolean logic that they aver arise from this identification. 
  The present work does capture this qualitative observation   more sharply and quantitatively, through  the emergence of  negative eigenvalue(s)  of the corresponding pseudo-projections.  
  
  
  Pseudo probability schemes
 capture non-classical probability  by violating the very first Kolmogorov axiom for probability theory which states that the probability for any event $E$, $P(E) \in \mathbb{R};~
P(E) \ge 0$. Some random variables can be assigned negative probabilities and, consequently, there will be other random variables for which the assignment of probabilities would exceed one (this can be seen, e.g., in the case of negation). The other two axioms are preserved. 

Quantum mechanics is distinguished from other theories which
employ negative probability \cite{Popescu94} by virtue of the fact that the number of negative probabilities is controlled strictly by the spectrum of pseudo projections. 

We devote the rest of the paper to illustrate the salient features of non-classicality for two level systems, with a  very brief mention of pure state entanglement. Only mutually non-commuting projections will be used in construction of pseudo projections, henceforth.

\section{Non-classicality in two level systems}
\label{s3}
The phase space underlying  a two level system is compact, and is  endowed with the Poisson bracket 
\newline
$\{\vec{S}\cdot\hat{m},\vec{S}\cdot\hat{n}\}\ = \vec{S}\cdot(\hat{m}\times \hat{n})$.
The quantum mechanical states have the form  $\rho =\frac{1}{2}(\mathbb{1}+\vec{\sigma}\cdot\vec{P}); |\vec{P}| \le 1$. Observables  have the form $A_i = \vec{\sigma}\cdot\vec{m}_i$, apart from the trivial identity. The two    eigen-projections belonging to the respective  eigenvalues $\pm\vert \vec{m}\vert$ of  $A$ are given by  $\pi_{a} = 
\frac{1}{2}(1+ a \vec{\sigma}\cdot\hat{m})$; $a=\pm 1$. 

The set  \{$\pi_a$\} exhausts all possible projections and, hence, outcomes of all possible measurements involving single observables. Thus, classically - in the sense of probability -- it should be  possible to assign joint  probabilities  for possible values $\{\pm 1\}^N$ with respect to sets of directions \{$\hat{m_i}\};~i=1,2,\cdots N$,  with the proviso
that  for each of them, 
\begin{equation}\label{jpg}
P_c\Big(A(\hat{m}) =1 \Big) = p_1 \implies  P_c\Big(A(-\hat{m}) = 1\Big) =  1-p_1.
\end{equation}
A complete  probability scheme would  then consist of assigning joint probabilities for all points on a unit sphere (see Eq.(\ref{jpg})), which would but be very highly correlated. If such a scheme exists for a given quantum mechanical state, 
the underlying classical space would be quite distinct from the phase space for the two level system which we started with. A correspondence between the two spaces exists only when $N\le 3$ and the directions are mutually orthogonal. 

Quantum mechanics forbids the existence of such probability assignments, even when $N=2$. This gets reflected in the emergence of negative entries in the corresponding schemes.

\subsection{Non-classicality with respect to pairs of observables}
\label{twoo}
Let $A_1, A_2$ be two observables, each having two outcomes:  $a_{1,2} = \pm 1$. The corresponding complete set of pseudo-projections for their joint occurrence is given, in terms of respective projections, by
\begin{equation}
\mathbf{\Pi}_{a_1 a_2}=\frac{1}{2}\{\pi_{a_1},\pi_{a_2}\};~ a_{1,2}=\pm 1.
\end{equation} 
The four pseudo-probabilities for a qubit in a state $\rho$ can be read off as 
\begin{equation} 
\label{PPQ}
{\cal P}_{a_1 a_2}
=  \frac{1}{4}\Big\{ 1 +a_1a_2\hat{m}_1 \cdot \hat{m}_2 + \vec{P}\cdot (a_1 \hat{m}_1 + a_2 \hat{m}_2)  \Big \}.
\end{equation}
As mentioned, the marginal scheme obtained by summing over  $a_1$ yields the quantum mechanical probabilities for $A_2$ and vice versa.

Several quick conclusions may be drawn from Eq.(\ref{PPQ}).  It follows from Eq.(\ref{PPQ}) that at most one entry in the pseudo probability scheme can be negative, while,  the requirement of proper marginals would allow for two negative values. This feature is characteristic of quantum mechanics, distinguishing it from other physical theories/models \cite{Popescu94}.

Significantly, if no condition on the two directions is imposed, all states, except the completely mixed state, turn out to be non-classical. More explicitly, so long as $\vert \vec{P}\vert \neq 0$, one can always find two observables for which the scheme becomes negative. Let us, therefore, impose the restriction 
 $\hat{m}_1\cdot\hat{m}_2 =0$, which  is equivalent to simultaneous specification of two orthogonal components of the classical spin vector. With this restriction,   the  condition on classicality gets diluted. Indeed, a state would be deemed to be classical  so long as  $\vert \vec{P}\vert \leq \frac{1}{\sqrt{2}}$, covering about 70\% of the volume of the state space. This conclusion, together with extension to three orthogonal observables, is dual to the POVM for joint measurement of observables\cite{Busch86}. 

Remarkably, Eq.(\ref{PPQ}) coincides with the expression  obtained by Cohen and Scully
 \cite{Cohen86} who, starting with  quantum analog of characteristic functions for a pair of observables in a two level system, computed  
the probability for their joint outcomes. This exact agreement merits further investigation. 
  More pertinently,  the  present work, apart from being much simpler,  clarifies the precise meaning of what Cohen and Scully \cite{Cohen86} mean by joint probability, and 
goes beyond, by developing a  general framework with applicability to any number of observables.

\subsection{Non-classicality with respect to triplets of observables}
\label{threeobv}
Let ${\bf \Pi}_{ab}$ be a pseudo-projection. It has an important physical interpretation. It  represents conditional probabilities which are inherent to quantum mechanics. For, if a system is in a state $\rho_a = \pi_a$,  then $Tr (\rho_a\mathbf{\Pi}_{ab}) \equiv P(b\vert a) =P(a\vert b)$ is but the Born rule for  the probability that the outcome of a measurement of $B$ yields a value $b$, and vice versa.
   Von Neumann observes  that the symmetry in the Born rule is a defining feature of quantum mechanics \cite{vonneumann55}.  

 A similar requirement on classical conditional probability would force the equality \cite{Accardi81}
\begin{equation}
P_c(a|b) \doteq \frac{P_c(a,b)}{P_c(b)}= P_c(b|a) \doteq \frac{P_c(a,b)}{P_c(a)}
\end{equation}
which holds if, and only if, both $P_c(a)$ and $P_c(b)$ are uniform distributions, and thus,  $P_c(a,b)$ is itself uniform.
 This result is in consonance with our finding that no joint probability description exists for two observables, unless the state is completely mixed. 
Is it possible that the uniform distributions also may retain non-classical features, not revealed by pairs of observables? To answer this, we consider sets   of three observables. 

Let $A_1$, $A_2$ and $A_3$ be three observables,  $A_i =\vec{\sigma}\cdot \hat{m}_i$. The corresponding pseudo projections, and  hence, the resultant pseudo-probability scheme is, however,  not unique. Associated with any  given joint outcome $\{a_1, a_2, a_3\}$, of the three observables,  there are  three distinct inequivalent  unit-pseudo-projections (the projections are mutually non-commuting),

\begin{eqnarray}
{\bf \Pi}^{(1)}_{a_1 a_2 a_3} = \frac{1}{2} (\pi_{a_1} \pi_{a_2} \pi_{a_3} + \pi_{a_3} \pi_{a_2} \pi_{a_1})  \nonumber\\
{\bf \Pi}^{(2)}_{a_1 a_2 a_3} = \frac{1}{2} (\pi_{a_3} \pi_{a_1} \pi_{a_2} + \pi_{a_2} \pi_{a_1} \pi_{a_3}) \nonumber \\
{\bf \Pi}^{(3)}_{a_1 a_2 a_3} = \frac{1}{2} (\pi_{a_2} \pi_{a_3} \pi_{a_1} + \pi_{a_1} \pi_{a_3} \pi_{a_2})
\end{eqnarray}

One can form further  convex combinations of the three unit pseudo projections. Each combination is a legitimate representative of the parent indicator function.
 All these constructions are mutually inequivalent.

Be it as it may, inequivalent pseudo projections can still exhibit a broad class-equivalence when their action on the full state space is considered. They may exhibit the same universal features. As an example, we may note that the three unit pseudo projections, being related to each other by permutations, possess this broad equivalence. This extends to other classes also provided the members are related to each other  by permutation symmetry.

Of all the representatives, the completely symmetric pseudo projection, obtained by Weyl ordering is of special interest (see \ref{cps}). We focus our attention to the scheme obtained by Weyl ordering. The  associated pseudo probabilities have the form
\begin{eqnarray}
\label{tpp}
{\cal P}_{a_1 a_2 a_3} & = & \frac{1}{8} \Big( 1 + \vec{P} \cdot 
\sum_{i=1}^3 a_i \hat{m}_i +  \sum_{i < j =1}^3 a_ia_j\hat{m}_i \cdot \hat{m}_j  \nonumber \\
& + & \frac{1}{3}  a_1a_2a_3\sum_{(ijk)}\vec{P} \cdot \hat{m}_i\hat{m}_{j} \cdot \hat{m}_{k} \Big)
\end{eqnarray}
where ($i,j,k$) are distinct and vary cyclically in the last term; $ a_i = \pm 1$.
 
Eq. (\ref{tpp}) answers the question raised at the beginning of this section. Even the completely mixed state is not immune to non-classical behaviour.  To see this, consider the symmetric coplanar geometry of the three directions given 
by $\hat{m}_i\cdot\hat{m_j} = -\frac{1}{2}; i \neq j$, and construct the scheme for the completely mixed state ($\vert \vec{P}\vert =0$).  We get, two  negative pseudo probabilities
${\cal P}_{111}= {\cal P}_{-1-1-1} = -\frac{1}{16}$, with the other six pseudo probabilities being positive and having equal values,  dashing any hope that the completely mixed state is classical with respect to any set of observables.

  
  We conclude the section by revisiting the restriction that the three observables be mutually orthogonal. For this case, which has received much attention in the context of POVM, the assignments of values $\pm 1$ is perfectly consistent with the classical requirement. The corresponding scheme is, however, non-negative only if
  $\vert \vec{P} \vert \leq \frac{1}{\sqrt{3}}$. This renders about 58\% of the volume classical.

\section{Quantitative features of non-classicality}
We now examine more quantitative features of non-classicality that emerge from pseudo projections. We begin by
designing  a suitable measure for non-classicality of  a state. We call this measure, negativity, and define it as

\begin{equation}
\mathcal{N} = \frac{1}{2}(\sum_i \vert \mathcal{P}_i \vert - 1)
\end{equation}
where the summation runs over all entries in the pseudo-probability scheme. A  state is  non-classical, with respect to a set of observables,  if $\mathcal{N} > 0$. For any given state, ${\cal N}$ is a  monotonically increasing  function of purity, though not necessarily strictly. Of greater interest is the relative volume of the state space that is rendered non-classical by a family of observables. As a concrete example we consider the set of all pairs of observables,
$\{\vec{\sigma}\cdot \hat{m}_1, \vec{\sigma}\cdot \hat{m}_2\}$. Then negativity is given, for the special geometry 
 $\vec{P} \parallel \hat{m}_1 + \hat{m}_2$ by 
 \begin{equation}
\label{neg1}
\mathcal{N} = \frac{1}{2} \Big( \vert \vec{P} \vert \cos \frac{\theta}{2} - \cos^2 \frac{\theta}{2} \Big).
\end{equation}
Here $\hat{m}_1 \cdot \hat{m}_2 = \cos \theta$.  It follows that the state is non classical 
 only so long as $\vert \vec{P} \vert > \cos \frac{\theta}{2}$.  The full behaviour of negativity is shown in 
 Fig.         \ref{fig:region_1} for two values of polarisation. More conclusions can be drawn from this figure: 
 
 \begin{figure}[!ht]
\includegraphics[width=\linewidth]{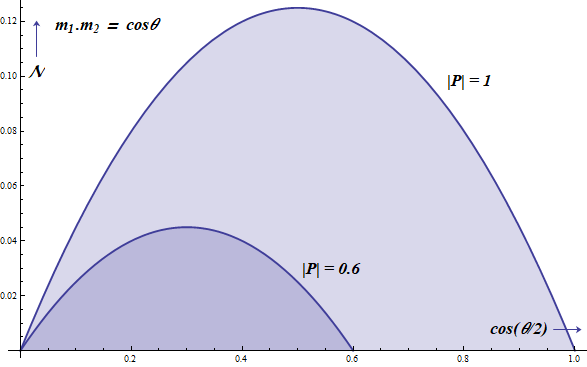}
\caption{Variation of negativity $\mathcal{N}$ with the geometry of the observables for different values of purity.}
\label{fig:region_1}
\end{figure}

\begin{enumerate}
\item For a fixed value of $\theta$, ${\cal N}$ is a strictly increasing function of purity. Thus, it can act as  a measure of coherence as well.

\item The maximum value of ${\cal N}$ is also a  strictly increasing monotonic function of purity; ${\cal N}_{max} = \frac{\vert \vec{P} \vert^2}{8}$.


\item The region of the parameter space (the relative volume in the space of observables) over which negativity is non-zero, shrinks as the mixedness of the state increases. For example for the completely mixed state $\mathcal{N}$ is zero for all possible combinations of observables whereas, for the pure state negativity is non-zero over the entire range of $\theta \in (0,\pi)$.
\end{enumerate}

\subsection{Negativity as a pure state entanglement monotone}
Finally, we round up the discussion with an application to quantify pure state entanglement in two qubit states.
The degree of entanglement in a pure two qubit state is uniquely determined by the mixedness  of it's reduced density matrix. Negativity, being a monotonic function of purity (of the reduced density matrices), can therefore be construed as a useful entanglement monotone for pure two qubit states. For instance,  the quantity $\mathcal{M} = 1 - \{\mathcal{N}_{max}(\rho_r)/\mathcal{N}_{max} (\rho_p) \}$ serves as a valid measure of entanglement for a pure two qubit state $\vert \psi \big>$; here $\rho_r$ is the reduced density matrix of $\vert \psi \big>$ while $\rho_p$ is a pure state belonging to the same subsystem and $\mathcal{N}_{max}$ is calculated form Eq.(\ref{neg1}). $\mathcal{M}$ goes to zero when $\vert \psi \big>$ is factorizable and to unity when the state is maximally entangled. It is easy to construct similar quantities from pseudo-probability schemes involving larger number of observables. Nevertheless the case involving just a pair of observables continues to remain the simplest.  Though this observation may appear to be rather simplistic, we show in the next paper that a judicious combination of logical propositions and combinations of pseudo probabilities permit construction of  a series of witnesses  for  entanglement in two qubit states.

\section{Conclusion}

In summary,  quantum probabilities and their underlying logic find their natural  formulation through our framework,  in the language of pseudo-projections and pseudo-probabilities.  This framework is completely operational and allows for a comprehensive description of non-classicality. This opens up a fertile field where non-classicality may be studied in all its avatars. In this paper we have focussed on application  to  two level systems and have developed a quantitative measure of non-classicality that can be used as a measure of coherence. We have further shown how this measure can be harnessed to infer the amount of entanglement in a pure two qubit state. Many interesting results  follow when we extend our work to multiparty systems as well as to higher dimensional systems.
These will be reported in subsequent publications.

\appendix
\section{Pseudo projections have at least one negative eigenvalue}
We show, in this appendix, that pseudo projections in any dimension $D$, and with any number of incompatible observables $N$,  have, at least,  one negative eigenvalue. 
Consider, first, two incompatible projections in $D=2$. The pseudo projection
\begin{equation}
{\bf \Pi}^{\hat{m}_1 \hat{m}_2}_{a_1a_2} = \frac{1}{2}\{\pi^{m_1}_{a_1}, \pi^{m_2}_{a_2}\}
\end{equation}
 has exactly one negative eigenvalue. If we now increase the number of observables, since ${\bf \Pi}^{\hat{m}_1 \hat{m}_2}_{a_1a_2}$ is the marginal of the  complete set of the corresponding pseudo projections, it follows that the result is valid for any value of $N$ as well.  We extend the result to $D\ge 3$, by induction. Indeed, a pseudo projection defined in $D\ge 3$ yields a pseudo projection in $D=2$ on taking its projection to the appropriate two dimensional subspace, thereby leading to a contradiction if the spectrum of the parent pseudo projection were to be non negative.

\section*{Acknowledgement}
Soumik and Sooryansh thank the Council for Scientific and Industrial Research (Grant no. - 09/086
(1203)/2014-EMR-I and 09/086 (1278)/2017-EMR-I) for funding their research. The authors thank Riddhi Ghosh and Ritu Rani for discussions.

\end{document}